\documentclass[preprint]{elsarticle}
\usepackage{lineno,hyperref}
\modulolinenumbers[5]

\journal{Journal of \LaTeX\ Templates}

\usepackage[utf8]{inputenc}
\usepackage{graphicx}	
\usepackage{amsmath}	
\usepackage{amssymb}	

\journal{New Astronomy}

\begin{document}

\begin{frontmatter}

\title{On the possible gamma-ray burst--gravitational wave association in GW150914}

\author{
Agnieszka Janiuk,$^{1}$ 
M. Bejger,$^{2}$
S. Charzyński$^{1,3}$
and P. Sukova$^{1}$}
\address{
$^{1}$ Center for Theoretical Physics,
Polish Academy of Sciences, Al. Lotników 32/46, 02-668 Warsaw, Poland \\
$^{2}$ Copernicus Astronomical Center,
Polish Academy of Sciences, Bartycka 18, 00-716 Warsaw, Poland \\
$^{3}$ Chair of Mathematical Methods in Physics, University of Warsaw, Pasteura 5, 02-093 Warsaw, Poland
}
\fntext[ead]{agnes@cft.edu.pl}

\begin{abstract}
Data from the Fermi Gamma-ray Burst Monitor satellite observatory suggested that the
recently discovered gravitational wave source, a pair of two coalescing black
holes, was related to a gamma-ray burst. The observed
high-energy electromagnetic radiation (above 50 keV) originated from a weak transient source
and lasted for about 1 second. Its localization is consistent with the
direction to GW150914. We speculate about the possible scenario for the formation
of a gamma-ray burst accompanied by the gravitational-wave signal. Our model invokes
a tight binary system consisting of a massive star and a black hole which leads
to the triggering of a collapse of the star's nucleus, the formation of a second black hole,
and finally to the binary black hole merger. For the most-likely configuration
of the binary spin vectors with respect to the orbital angular momentum in the GW150914 event,
the recoil speed (kick velocity) acquired by the final black
hole through gravitational wave emission is of the order of a few hundred km/s and this
might be sufficient to get it closer to the envelope of surrounding material and capture a small
fraction of matter from the remnant of the host star. The gamma-ray burst is
produced by the accretion of this remnant matter onto the final black
hole. The moderate spin of the final black hole suggests that the gamma-ray burst jet
is powered by weak neutrino emission rather than the Blandford-Znajek
mechanism, and hence explains the low power available for the observed GRB signal.
\end{abstract}

\begin{keyword}
black hole physics; accretion, accretion disks; gravitational
waves; neutrinos
\end{keyword}

\end{frontmatter}

\section{Introduction} \label{s:Introduction}

Gamma-ray bursts (GRBs) are extremely energetic, transient events observed
from all directions in the sky at high energies. Their known cosmological origin
requires the physical process that produces them to be a cosmic explosion of
great power. Proposed mechanisms involve the creation of a black hole (BH) in a
cataclysmic event. This may either result from the collapse of a massive
rotating star, or via the merger of two compact objects, e.g. binary neutron stars
or a BH and a neutron star. These two scenarios may produce long (\textgreater\;2
seconds) or short (\textless\;2 seconds) GRBs, respectively. The so-called `central engine' of this
process is composed of a hot and dense accretion disk with a hyper-Eddington
accretion rate (up to a few $M_\odot s^{-1}$) near a spinning BH and
fast relativistic jets that are launched along the BH's axis of rotation.
The angular momentum of the BH is usually invoked as a source of power of jets.
In addition, the annihilation of neutrino-antineutrino pairs produced in the
nuclear density plasma of the accretion disk can contribute to the jet power
(or powering jets). These fast jets produce GRB radiation that ultimately
can be observed far away from the central region.

These common scenarios may be related to gravitational wave (GW)
emission before and during the action of such an engine. Apart from the strong GW
emission produced by the time-varying mass quadrupole of an inspiraling and
merging compact binary system, several other suggestions were put forward, e.g.
neutrino-induced GWs \citep{2009PhRvD..80l3008S} or disk precession
\citep{2010A&A...524A...4R}. These, however, would be rather weak signals, most
probably below the sensitivity limits of current GW detectors.

The recent observation of a GW signal by the two Advanced LIGO detectors
on September 14, 2015 \citep{2016PhRvL.116f1102A} is related to
a merger of two BHs in a binary system. Both, the masses and spins of the
initial BHs and of the product of the merger were constrained from the
amplitude and phase evolution of the observed gravitational waveform.
In principle, mergers of binary BHs may be associated with
an electromagnetic emission (a GRB), if a sufficient supply of matter for the
accretion is involved at any stage of the merger, or after the GW event.
As hypothesized
in our previous work \citep{2013A&A...560A..25J}, a merger of a massive, rotating star
with a companion BH, in a system that evolved from a high mass X-ray
binary, may result in the collapse of the star's core. The merger of
the collapsed core, which is a newly formed BH, with its companion, would be then
the source of a transient emission seen in GWs. The accretion of matter onto
the core BH before the merger, and onto the final BH after the
merger, would be the source that powers the GRB. Potentially, either one or two
GRB signals could be observed, depending on the geometry of the system and
the observer's viewing angle. In the following, we elaborate on this scenario in
the context of a short duration, hard burst that could be associated with the GW150914 signal.

\section{Constraints for the GW150914 GRB} \label{s:Data}

The GRB that could be tentatively related to GW150914 by the observations
of the Fermi satellite had a duration of about 1 second and appeared about 0.4
seconds after the GW signal \citep{2016arXiv160203920C}. The two events were
temporally coincident and, within the limit of uncertainty of the two LIGO
interferometers and the Fermi detector capabilities to localize the GW source
and the electromagnetic source in the sky, could also be associated spatially.
The source of the GW was interpreted to be a merger of two BHs of the
masses of $36^{+5}_{-4}\ M_\odot$ and $29^{+4}_{-4}$
\citep{2016PhRvL.116f1102A}. The final BH parameters are estimated
to be of $62^{+4}_{-4}\ M_\odot$ and $0.67^{+0.05}_{-0.07}$ for its mass and
spin, respectively. The magnitudes and orientations of the spin vectors of the
two initial BHs are weakly constrained.
The probabilities that the angles between spins and the normal to the orbital
plane are between 45$^\circ$ and 135$^\circ$ are about 0.8 for each component
BH; spin magnitudes are constrained to be smaller than 0.7 and 0.8 at 90\%
probability, for the primary and the secondary BH, respectively. At the same
level of probability, the assumption of a strict co-alignment of spins with the
orbital angular momentum - a plausible astrophysical scenario in which
the BHs are produced from massive stellar progenitors - results in an upper limit of 0.2 and 0.3 for the
spins'  magnitude of the primary and the secondary BH, respectively (for more
details see Fig.~5 and related text in \cite{2016arXiv160203840T}). The
inferred posterior distribution of the GW150914
parameters disfavors an orientation of the total orbital angular momentum of
the system that is strongly misaligned to the line of sight (i.e., the system
was likely to be oriented face-on or face-away). Weak constraints on the magnitude and
the direction of the initial BH spins of GW150914 make it difficult to provide a
meaningful limit on the kick velocity of the resulting BH.

The event took place at a distance of
$410^{+160}_{-180}$ Mpc, corresponding to a redshift of about $z=0.09$
(assuming the standard cosmological model). The GRB fluence measured by Fermi
in the range 1 keV-10
MeV, is of $2.8 \times 10^{-7}$ erg cm$^{-2}$ which, for the distance inferred
from the GW observation, implies that the
source luminosity in gamma rays equals to $1.8^{+1.5}_{-1.0} \times 10^{49}$ erg/s.

\section{GRB+GW scenario} \label{s:method}

The scenario presented in \cite{2013A&A...560A..25J} describes the collapse of a
massive star followed by a binary BH merger. The progenitor, a massive
rotating star, is a member of a tight binary system with a companion BH. In order
to reconcile a coincidental GW and electromagnetic emission, we assume that after
the companion BH entered the star's envelope, the resulting interaction
with the stellar core causes its collapse into a second BH.
Additionally, some of the angular momentum will be stored in the envelope,
as it will be spun up by the transfer of the angular momentum from the
companion BH. As a natural consequence, rotationally supported torus is formed
in the equatorial plane.

\begin{figure}
\includegraphics[width=\columnwidth]{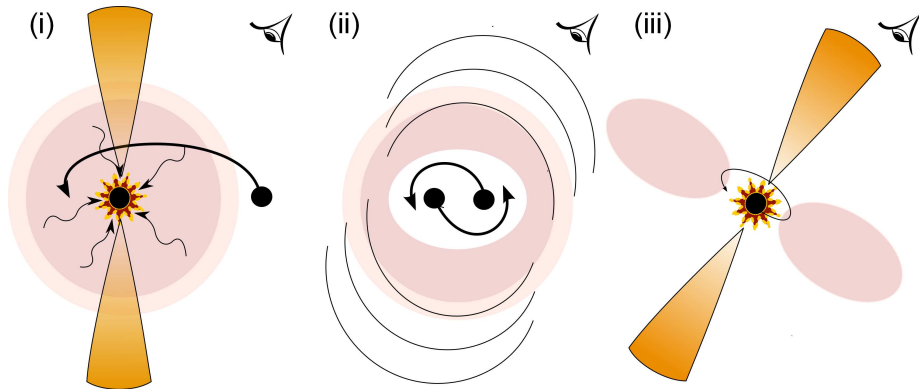}
\caption{Cartoon picture of the proposed scenario (stages i, ii, and iii are shown from left to right; see text). The BH in a binary
system induces core-collapse of the companion star (i).
Fallback accretion of
matter form the star's envelope might be accompanied by a weak jet, offset from
the line of sight. Next, binary BHs merge inside a circumbinary disk
(ii). As a result of the merger, the spin vector of the final BH changes its
direction. In the last phase, the remnant matter of the star's envelope is
accreted onto the final BH (iii).
}
\label{fig:cartoon}
\end{figure}

Our model involves three distinct stages of the binary evolution, namely (i) the core
collapse and accretion onto the core, (ii) the BH merger inside a
circumbinary disk in the interior of the collapsing star, and (iii) further
accretion of the remaining matter onto the final BH. Stages (i) and
(iii) may create favorable conditions to produce/release energies such as observed in jets,
and the resulting GRB signals; stage (ii) is the main engine
of the GW signal. The three stages are pictured in Fig.~\ref{fig:cartoon}.

Both GRB jet types, the one related to the progenitor's collapse and the other to
the accretion onto the final BH, may occur unnoticed to an observer if
the jets are collimated into narrow cones and the BH spins configuration favors
one specific line of sight. However, even if the axis of the first GRB is oriented
unfavorably towards the observer (i.e., offset with respect to the line of sight),
the second GRB which happens after the merger may be pointing towards the
observer. The direction of its axis should be coincident with the spin direction of
the final BH, which is the result of the two initial BH spins and the
evolution of the system, i.e., it may not be the same as the direction of the first
GRB.

In addition, the final BH may receive a natal kick, with a magnitude
depending on the BH mass ratio and the configuration of the initial BH spin vectors.
Therefore, stage (iii) may in principle lead to the GRB engine leaving its host site and approaching the inner edge of circumbinary disk.

The first phase, core-collapse, can be treated semi-analytically as in
\cite{janiuk08}. In that work, we considered two distinct cases:
a homologous accretion of
the envelope and a large increase of the subsequently created BH mass, or the accretion
through a torus, and wind outflow. The latter, if supported by a centrifugal
force and driven magnetically, can take away up to about 75\% of the mass
\citep{2013ApJ...776..105J} from the rotating torus.
Nevertheless, in the
current context, we suppose that no massive wind was associated with
GW150914, since the observations do not support a presence of large amounts of
mass in the vicinity of the GRB there. We also do not concentrate
here on the details of this phase, because the GRB was detected
{\it after} the GW signal. In the following, only these two phases are considered in detail.

\subsection{Black hole merger} \label{s:numerics}

We assume the BH merger occurred inside a circumbinary accretion disk
within the collapsing envelope of a massive star, and hence adopt the vacuum approximation
for the merger simulation. The merger is computed using the  \texttt{Einstein
Toolkit} computational package.\footnote{{\tt http://www.einsteintoolkit.org}}
The numerical methods used are based on finite difference computations on a
gridded mesh \citep{2012CQGra..29k5001L}. The technique of the adaptive mesh
refinement is used to cover a large volume with low resolution, and to cover
regions around BHs with a high-resolution grid. In our fiducial
simulation we use 7 levels of refinement (grid spacing differs by a
factor of $2^{6}$). The initial data contain the given masses, momenta and
spins for each BH and the initial separations of components. We
adopt the Cartesian grid with $48\times48\times48 M$, and resolution from $\Delta
x=2.0M$ for the coarsest grid to $\Delta x=0.03125M$ for the finest grid.

\begin{table*}
\begin{center}
\caption{Summary of the binary BH merger
 models. The initial separation of components is equal to $d$ and the initial momenta are $p$ and $-p$.
 The parameters $m_i$, $a_i$ and $\phi_i$ are the initial values of mass parameters, dimensionless spins and angles between spin directions and orbital angular momentum direction for each component. $M_1$, $M_2$ and $M_3$ are the total ADM masses of the components of binary BH and the mass of the product of the merger.
 The values are given in geometric units ($c=1$, $G=1$) and for the total ADM mass of the initial system set to 1, so in order to obtain values with proper units, separation, momenta and all the masses have to be multiplied by $GM/c^2$, $Mc$ and $M$ respectively, where $M$ is the initial mass of the BBH system (for GW150914, $M=65 M_\odot$).
 $a_3$ is the dimensionless spin parameter of the final BH, with $\phi_3$ being the angle between the direction of the spin and the direction of the orbital angular momentum of the initial system. The gravitational kick velocity of the final BH is equal to $v$.} \label{tab:par}
 \small
 \setlength{\tabcolsep}{3pt}
\begin{tabular}{|r|c|c|c|c|c|c|c|c||c|c||c|c|c|c|}
 \hline
 & \multicolumn{10}{|c||}{ Initial state} & \multicolumn{4}{|c|}{Final state}
\\
 \hline
 & \multicolumn{8}{|c||}{Puncture parameters}
 & \multicolumn{2}{|c||}{ADM}
 & \multicolumn{4}{|c|}{ADM values}
\\
 \hline
 \!run\! & $d$ & $p$ & $m_1$ & $m_2$ & $a_1$ & $ a_2$ & $\phi_1$ & $\phi_2$ & $M_1$ & $M_2$ & $M_3$ & $a_3$ & $\phi_3$ & \!$v\,[\frac{km}{s}]$\!\!
\\
 \hline
 0 & 10 & 0.093 & 0.541 & 0.443 & 0 & 0 & $0^\circ$ & $0^\circ$ &
 0.555 & 0.457 & 0.96 & 0.68 & $0^\circ$ & $120$
\\
 \hline
 1 & 10 & 0.093 & 0.53 & 0.432 & 0.2 & 0.3 & $0^\circ$ & $0^\circ$ &
 0.552 & 0.46 & 0.96 & 0.76 & $0^\circ$ & $130$
\\
 \hline
 2 & 10 & 0.093 & 0.53 & 0.432 & 0.2 & 0.3 & $10^\circ$ & $0^\circ$ &
 0.552 & 0.46 & 0.96 & 0.76 & $0.5^\circ$ & $130$
\\
 \hline
 3 & 6 & 0.139 & 0.52 & 0.424 & 0.2 & 0.3 & $10^\circ$ & $0^\circ$ &
 0.552 & 0.463 & 0.96 & 0.77 & $0.6^\circ$ & $100$
\\
 \hline
 4 & 6 & 0.139 & 0.507 & 0.409 & 0.3 & 0.45 & $10^\circ$ & $0^\circ$ &
 0.55 & 0.467 & 0.96 & 0.81 & $0.8^\circ$ & $280$
\\
 \hline
 5 & 6 & 0.139 & 0.484 & 0.385 & 0.4 & 0.6 & $10^\circ$ & $0^\circ$ &
 0.545 & 0.471 & 0.95 & 0.85 & $1^\circ$ & $200$
\\
 \hline
 6 & 6 & 0.138 & 0.415 & 0.341 & 0.7 & 0.7 & $90^\circ$ & \!\!$120^\circ$\!\! &
 0.557 & 0.459 & 0.97 & 0.72 & $25^\circ$ & $700$
\\
 \hline
\end{tabular}
\end{center}
\end{table*}

Since BBH mergers simulations are scalable with respect to the total mass of the system in all computations the total mass of the system is set to 1 for simplicity. As the total mass we mean the mass measured by a distant observer, namely the Arnowitt-Deser-Misner (ADM) mass \citep{1959PhRv..116.1322A}. Therefore, all the masses given in the table are simply fractions of the total mass of the initial BBH system.
The mass ratio of the merging components is 0.82 (ratio of the most-probable
values of estimated masses for GW150914).
The third order PN approximation is used \citep{PhysRevD.77.024027} to
find the initial momenta of the components of the binary on quasi-circular
orbits.
The product of the merger is described by its mass and spin, but we
also estimate the gravitational recoil speed from the analysis of
the linear momentum carried by the outgoing gravitational radiation
through the sphere of radius $42M$. We use the Weyl
scalar $\Psi_4$ multipole decomposition method (up to the order $l=4$) described in
\cite{Alcubierre:2008it}. The initial parameters and the
results of the simulations are presented in the Table \ref{tab:par}.

\begin{figure}[bbh]
\includegraphics[width=\columnwidth]{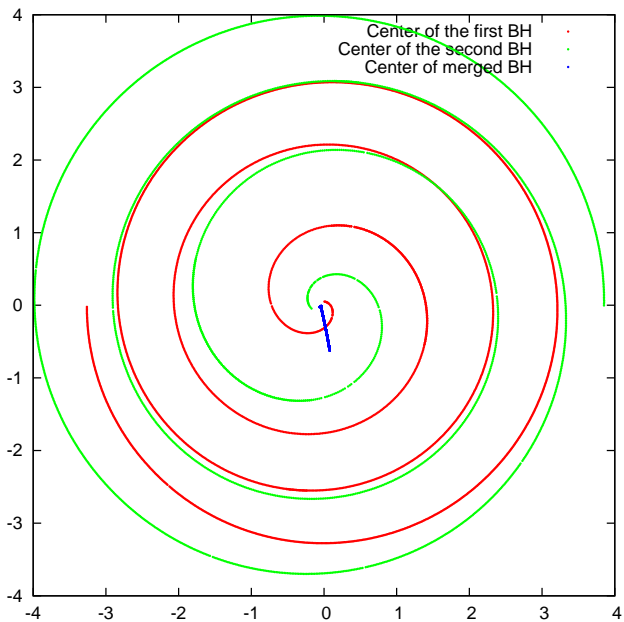}
\caption{
Last three orbits of the merging BHs (run 2 from the Table
\ref{tab:par}).}
\label{fig:orbits}
\end{figure}

\begin{figure}
\includegraphics[width=\columnwidth]{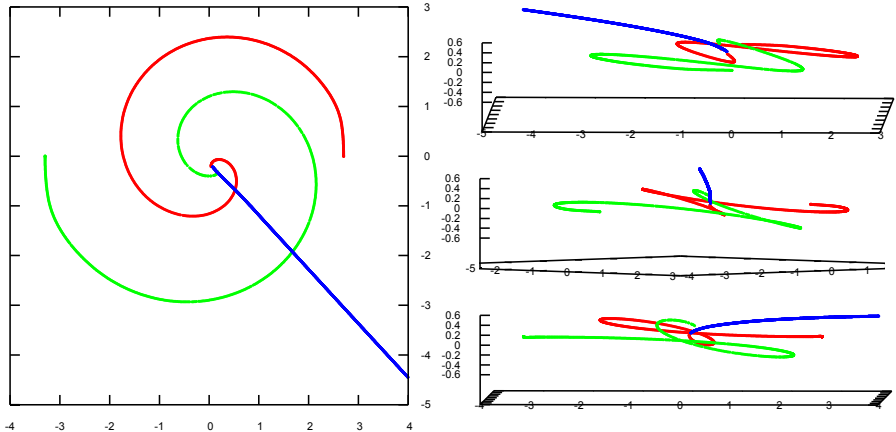}
\caption{Simulated orbits of the merging BHs of run 6 from the Table \ref{tab:par}.
Left: projection onto $xy$-plane. Right: three views from different directions almost perpendicular to the $z$-axis, showing the non planarity of orbits for this spin configuration.}
\label{fig:orbits3d}
\end{figure}

\begin{figure}
\includegraphics[width=\columnwidth]{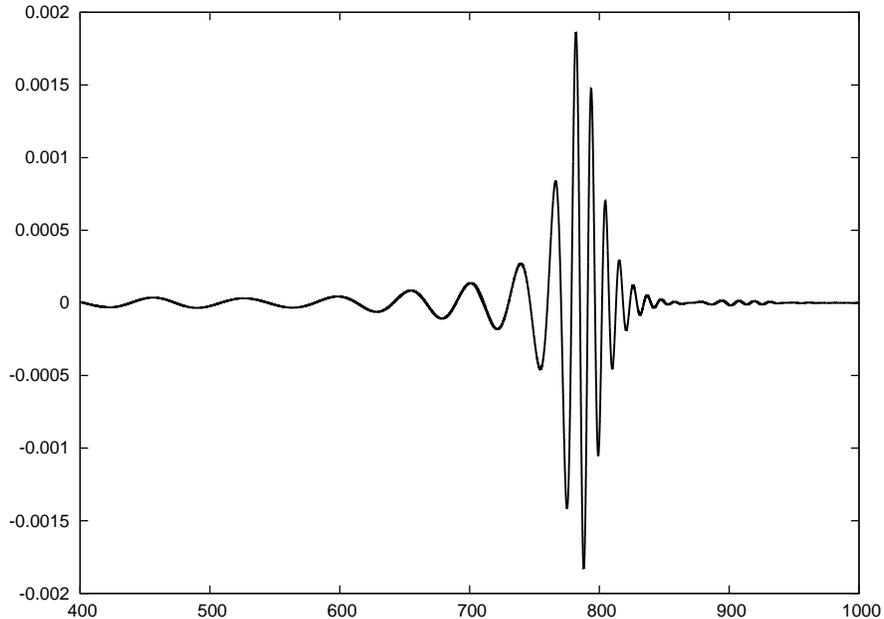}
\caption{Exemplary gravitational wave signal: the real part of the $l=2$, $m=2$ multipole component of the Weyl scalar $\psi_4$ extracted at the sphere of radius $42M$ for the run 2 in the Table \ref{tab:par}.}
\label{fig:radiation}
\end{figure}

The simulations were preformed for
a range of values for spin magnitudes and orientations consistent with
the (weak)
parameter estimation for the GW150914 signal \citep{2016arXiv160203840T}.
In general we assume that the spin vectors may be mildly
misaligned with respect to the orbital spin vector, a situation
that most likely occurs in the massive progenitor binary scenario.
Within this setup only the second GRB, which is related to
the jet produced by the accretion onto the final BH, would
be visible to the observer. We also simulate one case with
a strong misalignment of spins (case 6 in Table \ref{tab:par})
that results in a substantially higher recoil speed.
Fig.~\ref{fig:orbits} and~\ref{fig:orbits3d} show
a few exemplary orbits of binary BH. The extracted gravitational wave
signal for one of the simulations
is plotted on Fig.~\ref{fig:radiation}.

\subsection{Accretion onto a final black hole} \label{sec:harm}

The numerical method used for the computation of the GRB engine and an estimation
of its jet power is based on the axisymmetric, general relativistic MHD
simulation with the code {\it HARM-2D}, whose basic version was described by
\cite{2003ApJ...589..444G}. It uses a conservative, shock-capturing
scheme, and provides a solver for the continuity and energy-momentum conservation equations, assuming a force-free approximation.
This scheme was used recently for the studies of magnetized, radiatively inefficient
accretion flows in the Kerr black hole field \citep{2012MNRAS.423.3083M}.
Here we use our own numerical routines to
compute the cooling by neutrinos, as was described in
detail in \cite{2013ApJ...776..105J}.
The neutrino cooling processes adopted in our computations are
the reactions of electron and positron capture on nucleons, the electron-positron pair annihilation, nucleon bremsstrahlung and plasmon decay.
The leptons and baryons are relativistic and may have an arbitrary
degeneracy level, so that we compute the gas pressure using the appropriate Fermi-Dirac integrals. In the total pressure, we include also
the contributions from the free nucleons, pairs, radiation, trapped neutrinos, and Helium nuclei.

In contrast to the simplified method used in the dynamical computations
that we presented in \cite{2013ApJ...776..105J},
where the neutrino cooling rate
was used to update in every time step only the internal energy in the plasma,
in our current version of the HARM-2D code, a numerically
computed equation of state is used
throughout the computations. Self-consistently, the pressure is computed
as a function of density and temperature,
which in this case is not given by a simple
adiabatic relation, but tabulated.
We use the tables that store the internal energy, pressure,
and neutrino cooling rate, computed
as a function of temperature and density in the ranges
between $10^{2}-10^{14}$ K, and $10^{2}-10^{13}$g cm$^{-3}$, respectively,
and are logarithmically spaced and have $256\times 256$ grid points.
The EOS is therefore deeply
incorporated into the dynamical scheme, where we solve for the
inversion scheme between the so called 'primitive' (rest mass density, internal energy) and 'conserved' (momentum, energy density)
variables at every time-step (see e.g. \cite{2006ApJ...641..626N}), which is done by a multi-dimensional
Newton-Raphson routine.

Our initial conditions for the accretion flow are given by the
equilibrium torus solution, defined as in \cite{1976ApJ...207..962F}.
We use the grid resolution of $256\times 256$ cells in the $r$ and $\theta$ directions,
and the grid is spaced logarithmically in radius and concentrated towards the
equatorial plane.
To speed up the computations, our version of this code was
parallelized using the MPI
technique. The adopted physical parameters, i.e. the BH mass and angular momentum,
and the torus mass defined by the radius of the pressure maximum, are
supplemented by the initial geometry and strength of the magnetic field.
The latter, in our fiducial computations,
 is adopted as a standard, poloidal field given by the
$\phi$-component of its vector potential scaling with the density and
the initial $\beta=P_{\rm gas}/P_{\rm mag}=50$ (see e.g. \cite{2012MNRAS.423.3083M} for the discussion of various field configurations).
The resulting observables, the
neutrino emissivities, are computed with good accuracy, and we compare the resulting power from the integrated neutrino luminosity with that available
via the magnetic flux dragged through the black hole horizon (we note here that due to the limitations of our 2-dimensional scheme,
it is only an order-of-magnitude estimate).

\section{Results} \label{s:results}

From the computation of the binary BH merger, we extract the values of mass, spin and recoil speed of the merger product for given initial configurations of the binary components. Since
the GW150914 event observation didn't lead to the estimation of the recoil speed
for the final BH, we have performed a set of simulations with mass ratio
and spins values consistent with the estimated GW150914 parameters
\citep{2016arXiv160203840T}. Using this range of parameters,
one may estimate the upper limit on the
recoil speed for this event. The results of our simulations are gathered in Table \ref{tab:par}. The simulations confirm that the spin of the merger product is almost exactly aligned (with difference less than $1^\circ$) with the orbital angular momentum of the binary BHs for scenarios with more massive components a spin misalignment equal to $10^\circ$ for a large range of spin values. A $10^\circ$ change in the direction of the spin is enough for one of the GRB's to become unobservable.

For the eventually observed GRB event, we performed a numerical simulation
of the accretion of remnant matter onto the final BH. The parameters for
this fiducial model were as follows: the BH mass $M_{\rm BH} = 62 M_{\odot}$ and
its spin values $a=0.6$, $0.7$ and $0.8$, corresponding to the values inferred for the GW150914 event.
The mass of the accreting torus is not constrained by the LIGO data. It is therefore a free
parameter in our model, and we determine it using an appropriate density scale.
This scaling determines then the conditions in the torus for the nuclear
reactions to take place, in which the neutrinos and anti-neutrinos of three
flavors will be produced \citep{2013ApJ...776..105J}.

We investigated the two following setups.

\begin{figure}
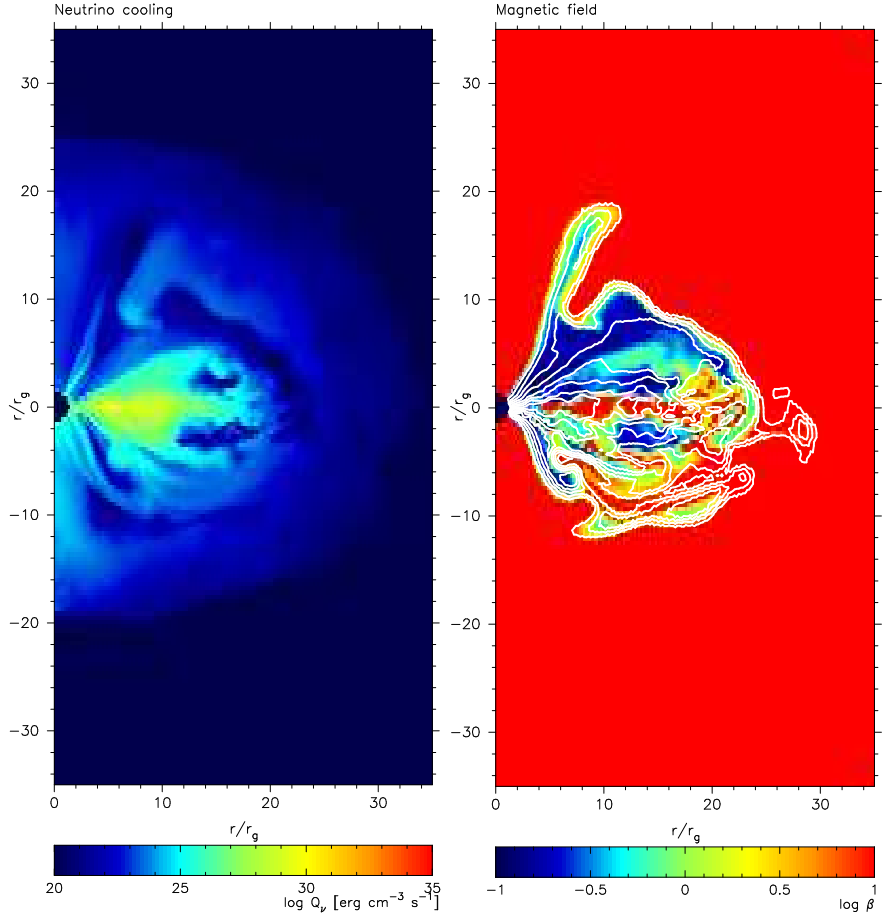

\includegraphics[width=\columnwidth,angle=270]{{map.neu.361.d200_r30}.ps}
\includegraphics[width=\columnwidth,angle=270]{{map.beta.361.d200_r30}.ps}
\caption{Results from the GR MHD simulation of a remnant torus accreting onto a BH in the GRB engine. Maps show the neutrino emissivity (left) and magnetic field lines topology together with the gas pressure to magnetic pressure ratio (right).  The parameters of the model are $a=0.6$, $M_{\rm BH} = 62 M_{\odot}$, $M_{\rm torus}\approx 15 M_{\odot}$. These snapshots were taken at time t=2000 M ($\approx 0.6$ s). The accretion rate through the torus at the inner boundary is about $5.56 M_{\odot}/s$.}
\label{fig:harm}
\end{figure}

\begin{figure}
\includegraphics[width=\columnwidth]{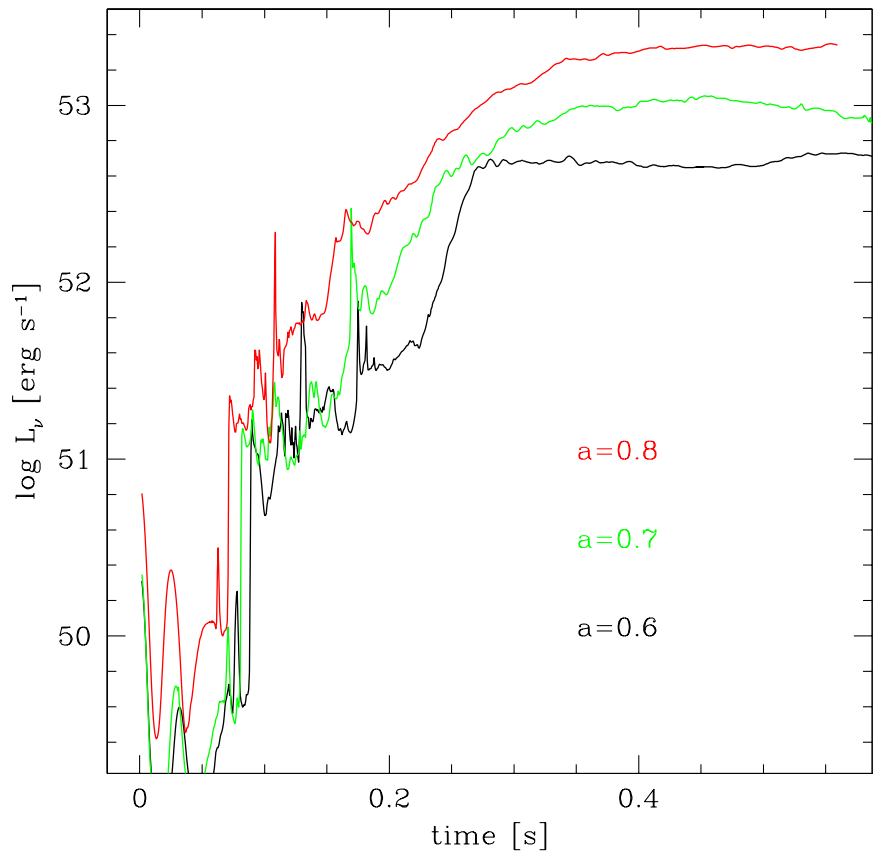}
\caption{Results from the GR MHD simulation of a remnant torus accretion.
The neutrino luminosity is plotted versus time, until $t=2000 M$.
Parameters of the model are $M_{\rm BH} = 62 M_{\odot}$, $M_{\rm torus}\approx 15 M_{\odot}$. The lines show models with BH spins $a=0.8, 0.7$, and $0.6$, from top to bottom.}
\label{fig:lnu}
\end{figure}

In Fig.~\ref{fig:harm} we show the neutrino emissivity, as computed from
the density and temperature distribution in the torus, in units of
erg s$^{-1}$ cm$^{-3}$, as well as the structure of the magnetic field. The results were taken
at the end of the simulation, at $t=2000 M$, which corresponds to about
0.61 seconds of real time for the assumed BH mass. Parameters of this
model were $a=0.6$, $M_{\rm BH} = 62 M_{\odot}$, $ M_{\rm torus}\approx 15
M_{\odot}$. The computed luminosity  emitted  in neutrinos is in this case
about $L_{\nu}=5\times 10^{52}$ erg/s at $t=2000 M$.
The total neutrino luminosity for the models with $a=0.7$, and $a=0.8$
was on average about 3 and 6 times higher, respectively,
than that for a fiducial model
(see Fig.~\ref{fig:lnu}).

For comparison, we also tested the case where the mass of the
torus is approximately equal to the final BH mass,
$M_{\rm torus} \approx 57 M_{\odot}$. We computed the total neutrino
luminosity, integrated over the emitting volume, at time $t=380$ M of the dynamical simulation,
 to be $L_{\nu}= 5 \times 10^{55}$ erg/s.
This
value in the dynamical simulation would increase even further, at t=2000 M it would be larger by $\sim 1.5$ orders of magnitude.
Therefore, the neutrino emission from such a massive torus
would exceed the neutrino luminosity
determined for the observed GRB \citep{2016arXiv160208436M},
by many orders of magnitude, and we conclude that this setup is
not realistic for the observed limits, regardless of the details
of the GRB power supply by neutrino annihilation and
jet production efficiency (see Discussion).

The power and luminosity available through the
Blandford-Znajek process \citep{1977MNRAS.179..433B}
in the present model is completely negligible
because of a too low value of the BH spin and no magnetization
of the gas at the horizon and in polar regions as
shown in the map in Fig.~\ref{fig:harm}.
We checked the magnetization value on the horizon,
$\beta(r_{\rm in})=B^{2}/\rho(r_{\rm in})$, where $\rho$ is the rest mass
density of the gas. The electromagnetic stress tensor is given by
\begin{equation}
T^{\mu \nu}_{\rm EM} = b^{2}u^{\mu} u^{\nu} + \frac{b^{2}}{2} g^{\mu \nu} - b^{\mu} b^{\nu},
\end{equation}
where $b^{\mu}$ is the magnetic four-vector, with $b^{t}=g_{i \mu}B^{i}u^{\mu}$ and $b^{i}=(B^{i}+u^{i}b^{t})/u^{t}$ and $u^{\mu}$ is the four-velocity. We compute the radial electromagnetic flux through the horizon
\citep{2004ApJ...611..977M}
\begin{equation}
\dot E = 2\pi \int_{0}^{\pi} d \theta \sqrt{-g} F_{\rm EM},
\end{equation}
where $F_{\rm EM} = - T^{r}_{t}$ and $g$ is the metric determinant.
In our models, we made computations
under the assumption of a weakly magnetized plasma,
with $\beta_{\rm init}=P_{\rm gas}/P_{\rm mag}=50$.
At the end of the simulations (at $t=2000 M$) the power transferred to the polar
regions of the BH via the Blandford-Znajek process was
$L_{\rm BZ} \equiv \dot E \approx 1.1 \cdot 10^{50}$ erg/s for $a=0.8$,
and $3.7 \cdot 10^{51}$ erg/s for $a=0.9$,
thus was giving a much smaller power to the GRB jets than the neutrinos.
Moreover, for the case of $a=0.6$, there was no net magnetic flux dragged
through the horizon in our simulation, so the BZ power was virtually zero.

We note here also, that a moderate value of the BH spin
affects the topology of the magnetic fields, so that they remain confined
to the torus plasma. The magnetically driven winds are therefore hardly launched. This
may also be the reason for quite
a low total neutrino luminosity, since not many neutrinos
are produced in hot, massive winds.

In Fig.~\ref{fig:lnu} we show the neutrino luminosity of the GRB engine,
presented as the
averaged neutrino
emissivity integrated over the simulation volume, as a function
of time. Initial conditions, which are based on the adopted pressure
equilibrium torus solution, are evolved, and after about $\sim 1000 M$ the
configuration reaches its dynamical shape. The neutrino luminosity is at its maximum then and will continuously decrease with time afterwards. (Note that
in the plot we use the physical units, with $t=GM/c^{3}$ and scaling for $M_{\rm BH} = 62 M_{\odot}$, so that $1000 M=0.304$ s.)

\section{Discussion} \label{s:discussion}

The GW150914 observation is just a first example from the incoming population of
binary BH mergers to be expected in the near future from the Advanced
LIGO and Advanced Virgo experiments \cite{2016arXiv160203842A}. Some of
them {\it may be} coincident with the electromagnetic observations.

By analyzing the properties of the merger one can easily compute
the mass and spin of the final BH
\citep{2005CQGra..22..425P,2006PhRvL..96k1101C,2006ApJ...653L..93B}. However,
the estimation of its kick velocity requires the evaluation of the linear
momentum carried away by the gravitational radiation during the merger
\citep{Alcubierre:2008it}, which in turn depends on a precise knowledge
of the binary components spins. For specifically chosen mass
ratios and spin configurations of the initial BHs, the kick velocity may exceed
4000 km/s, but such large recoils are very rare
\cite{2011PhRvL.107w1102L}.

There are hints for evidence for spin flips and past merger events 
\citep{1995ApJ...439...98C, 2003MNRAS.340..411L, 2009ApJ...697.1621G, 2012ApJS..201...23E}.
The theoretical effort is thus supported and motivated by observational discoveries like
the recent one which is a tentative detection of a GRB in coincidence
with the GW signal. If true, this would be the most spectacular finding.

In this work we propose a scenario that plausibly explains an `exotic'
GRB progenitor. We hypothesize that a system that contains a massive star and a companion BH orbiting
inside its envelope, triggers the core of the companion star to collapse due to the interaction with
the BH. We calculated the values of masses and spins of the binary components and the
merger product.
For the masses of GW150914 and a selected range of spin magnitudes and orientations,
we obtain a range of recoil speeds of $100-700$ km/s, where
the larger value correspond to the simulation with strongly misaligned spins.
With such a velocity, the recoiled BH can move by a distance
that may be significant for an accretion process in the last stage.
To see this, we employ the work of \cite{2012MNRAS.427.2660K} who studied the circumbinary disk with a cavity in the middle around which the mass piles up. The radius of the cavity in the equal mass case (mass ratio $q=1$) is estimated as $r_s + r_H$, with $r_s$ denoting the binary separation and $r_H$ the Hill radius, $r_H = (q/3)^{1/3} r_s$.
For a binary separation $r_s=5M$ just before the merger the cavity radius would be approximately equal $8.5M \simeq 750$ km. During the merger the cavity shrinks, with the violent movement of masses generating the GWs possibly disturbing its inner edge.
It is then likely that fast accretion of the accumulated matter onto the final BH is triggered while it moves towards the disk after the merger.

Additionally, the final BH may in principle capture some of the surrounding matter.
The amount of gas that is gravitationally bound to
it is determined by the recoil speed. The accretion power and duration available
to feed the GRB after the merger are
determined by the mass of the disk. Moreover, the outer parts of such a mini-disk
may be large enough to exhibit some misalignment from the plane perpendicular
to the BH spin vector. In such a case, disk and jet precession may
occur which would give an additional, periodic signal on top of the GRB emission. Possibly, an orphan afterglow signal in the lower
electromagnetic energy bands may be present.

\cite{2016arXiv160300511W} discussed
a progenitor scenario for GW150914 that involves
the core-collapse of a single,
chemically homogeneous, rapidly rotating, massive (mass of about $150 M_{\odot}$), single star.
This setup would produce a BH promptly, but
because of a negligible mass loss in a homogeneous star, the jet breaking
through the thick envelope would result in a GRB emission delayed by more
than 10 seconds with respect to the GW signal.
It should result from the Kerr parameter of the collapsing core
being significantly larger than unity, so the angular momentum of the newly born BH is lost via gravitational wave emission.
However, \cite{2016arXiv160300511W} does not show if the waveforms emitted during such a process
are compatible with the ones observed by LIGO. Should this scenario be correct, then also all the parameters of the GW progenitor estimated by \cite{2016PhRvL.116f1102A} need to be revised.
An alternative scenario of a binary system that consists
of two massive stars which undergo core collapses in the common envelope phase is more plausible, because in that case the GRB coincides with GW emission instantaneously. The two massive stars with an initial separation on the order of 1 AU would undergo core collapses one after another and experience twice the common envelope phase.
Similarly, in \cite{2016ApJ...821L..18P}, the authors discuss the possibility that the two massive stars evolve in a binary system and explode as core-collapse supernovae one after another. The matter from the envelope of the second supernova remains bound and finally accretes onto the BH after the merger, to power a GRB (see also \cite{2016ApJ...822L...9M} for the discussion of the properties of such a mini-disk).

Our scenario is in line with the second one proposed by \cite{2016arXiv160300511W}, as it involves the last stage of the binary system evolution.
Here, a binary consisting of a Wolf-Rayet star and a massive BH evolves out of an ultraluminous X-ray source phase,
and the BH is brought into the common envelope,
triggers the core-collapse and merges with the newly-formed BH that originated from the imploding helium core.
The timescale from the primary BH formation until the merger
does not have to be a typical one for the common envelope phase, and
depends significantly on the mass of the star \citep{1996A&A...310..489L}.
For a supergiant star, this timescale might be
of the order of months or shorter,
similar to a supermassive Thorne-Zytkow object with a core neutron star
\citep{1977ApJ...212..832T}, which can lead to the observable soft gamma ray repeaters and anomalous X-ray pulsars
(e.g. \cite{2015ApJ...799..233L}).

Another possibility, envisaged by \cite{2016ApJ...819L..21L},
is somewhat similar to both of the above mentioned and our scenario, as
the two BHs also
merge within a common envelope of a
very massive star. According to their model, two BHs must have formed simultaneously
from the two clumps that were created via the bar instability during the core collapse. While this scenario offers a possibility to explain the origin of the two BH masses, it also introduces a large uncertainty of the very process of such a non-axisymmetric
core collapse. In our scenario, the core collapse leads to the formation of
only one BH in a process which is assumed to be induced by the presence of a
companion BH in the common envelope. Additional transfer of angular momentum by the inspiraling BH into the envelope leads to an increase of the star's angular momentum and the formation of a circumbinary disk/accretion torus important for the subsequent GRB.

The accretion of the magnetized, rotating torus onto the BH is a commonly accepted
mechanism of jet launching in GRBs. The issue of the jet break through the
star's ejected envelope is another problem that all the `collapsar' studies
must take into account. In particular, the timescale for the jet break depends
on the details on the star's initial composition, its metallicity, rotation
profile and mixing effects. We argue that the motion of the companion BH
through the star and the respective angular momentum transfer cause the
disruption of the outer parts, so that a significant part of the mass is
expelled. In addition, the creation of the high-angular momentum accretion torus
in the inner part of the progenitor
helps to evacuate matter from the polar regions, which in turn allows the jet to
break easier: if the spin of the merger product is not strongly misaligned with
the binary orbital angular momentum, we expect that the jet would not be
significantly held back by the envelope. The maximum of the neutrino luminosity
obtained in our simulations is reached about 0.4 seconds after an equilibrium
torus, prescribed by our initial conditions, had formed. This may tentatively
give the lower limit for the
timescale when the jet appears after the BH merger. However, the jet sustains only as long as the torus material is consumed by the BH, so for about 3 seconds. Within this time, the neutrino-antineutrino
pairs must reach the polar regions and provide the energy for efficient jet acceleration to the high Lorentz factors, $\Gamma \sim 100$, so that the kinetic energy of the jet is converted ultimately to gamma rays.

The numerical studies of the accretion of magnetized, neutrino cooled matter onto a BH presented here are aimed to quantitatively estimate the conditions for the observed low luminosity of the resulting GRB.
As discussed by \cite{2016arXiv160208436M}, the upper limit for the total
isotropic equivalent of the luminosity, $E_{\gamma, \rm iso}$, from this GRB is at most $3\times
10^{52}$ erg/s, as was restricted by the non-detection of neutrinos by the
IceCube experiment. Generally, neutrino annihilation would lead to the GRB
luminosity of
\begin{equation}
L_{\nu, \bar{\nu}} = (1+z) E_{\gamma, \rm iso}(1-\cos \theta_{\rm j})/T_{90},
\end{equation}
where $\theta_{j}$ is the opening angle of the jet.
As derived in the numerical simulations by
\cite{2011MNRAS.410.2302Z} (see also \cite{2015ApJ...806...58L} for a
fitting formul\ae\ in a simpler steady-state 1-dimensional model of an NDAF
disk), the $L_{\nu, \bar{\nu}}$ luminosity scales with the BH mass,
spin and global average accretion rate to the power of $-3/2$, $-4/8$, and
$9/4$, respectively.  The efficiency of the neutrino energy deposition outside
the BH horizon, $\epsilon=L_{\nu, \bar{\nu}}/L_{\nu} \approx 0.05
\dot m^{5/4}$, and depends strongly on the BH spin. The accretion rate
of the order of $M_\odot/s$ will lead to the luminosity of the
explosion of the order of the canonical value for collapsars, i.e.,
$10^{51}$ erg/s, for a high spin of $a=0.95$ (and for a BH
mass of 3 $M_{\odot}$). For a non-rotating BH, this luminosity would
be obtained if the mass accretion rate were ten times larger.

The mass accretion rate is one of the uncertainties of this model. The simplest assumption
is that the mass of the accretion torus is of the same order as the mass of the
final BH, but it doesn't have to be the case. The neutrino luminosity
produced under the assumption that the torus mass is of about only $15
M_{\odot}$ fits better to the inferred upper limits, and does not require
any additional fine-tuning of neither the annihilation efficiency (which might be very
much different in case of magnetized disks than in simple NDAF models), nor the jets opening angle.

The GRB luminosity inferred from our simulations
can be reconciled with the observational
upper limits, for moderate spins of the final BH ($a=0.6-0.8$).
Furthermore, we can assume that not more than 10 per cent of the
electron-positron pairs that were created by the neutrino annihilation
contributed to the GRB fireball, and the rest might have fallen back into
the BH.
Finally, because the total
event lasted for $T_{90} = 1 s$, the large mass of the torus
and the mass accretion rate is not consistent with an estimate of
$T_{90}$.  Nevertheless, our
simulation shown in Fig.~\ref{fig:harm} is roughly consistent with the
observed
GRB duration and gives $T_{90} \approx M/\dot M \approx 15/5.5=2.7$ s.
In this case however, the longer-lasting `tail' of the GRB signal,
might rather have been below the
detection noise level.

We note here that the Fermi detection of a GRB
coincident with GW150914 has not been confirmed by deep Integral observations \citep{2016ApJ...820L..36S}.
However,
if this connection is real, or if in the future more coincident observations of
gravitational signals with GRBs will be seen, then we have to face the fact,
that there are mergers of two massive BHs in an environment
with a sufficient amount of matter to produce a GRB with short time delay
after the merger. Our computations show a possible mechanism for such a GRB
to emerge with the parameters given by the observed signal (masses and spins of the BH), which is based on neutrino cooling. Our scenario for
such a configuration is speculative, but it can be tested and verified
by further observations and simulations.

In conclusion, we propose and numerically compute a plausible model for the
GW emission being coincident in time with a weak GRB observed by Fermi.
To know whether this model is indeed realized in nature,
further searches for gravitational wave sources and their
electromagnetic counterparts are now essential, with both novel
experimental techniques and theoretical efforts in numerical relativity.

\section*{Acknowledgments}
We acknowledge comments and interesting discussions with Tomasz Bulik, Odele Straub and
Lukasz Wyrzykowski.
This work was supported in part by the grants no. DEC-2012/05/E/ST9/03914 and
UMO-2014/14/M/ST9/00707 from the Polish National Science Center.
We also acknowledge support from the Interdisciplinary Center for Mathematical Modeling of the Warsaw University, through the computational grant G53-5.

\bibliography{ptapap_ajaniuk_na} 


\end{document}